\documentclass[twocolumn,showpacs,superscriptaddress,preprintnumbers,amsmath,amssymb,prc]{revtex4}

\usepackage{graphicx}
\usepackage{dcolumn}
\usepackage{bm}
\usepackage{float}
\usepackage{color}
\usepackage{CJK}
\usepackage[colorlinks,linkcolor=blue,urlcolor=blue,citecolor=blue]{hyperref}

\begin{document}
\begin{CJK*}{UTF8}{gbsn}

\title{Magnetic field effects in peripheral heavy-ion collisions around 1 GeV/nucleon}
\author{X. G. Deng (邓先概)}
\affiliation{Key Laboratory of Nuclear Physics and Ion-beam Application (MOE), Institute of Modern Physics, Fudan University, Shanghai 200433, China}
\author{Y. G. Ma (马余刚)\footnote{Corresponding author: mayugang@fudan.edu.cn}}
\affiliation{Key Laboratory of Nuclear Physics and Ion-beam Application (MOE), Institute of Modern Physics, Fudan University, Shanghai 200433, China}
\affiliation{Shanghai Institute of Applied Physics, Chinese Academy of Sciences, Shanghai 201800, China}

\date{\today}

\begin{abstract}
Magnetic field effects on free nucleons are studied in peripheral collisions of $^{197}$Au + $^{197}$Au at energies ranging from 600 to 1500 MeV/nucleon by utilizing an isospin-dependent quantum molecular dynamics (IQMD) model. With the help of angular distributions and two-particle angular correlators, the magnetic field effect at an impact parameter of 11 fm is found to be more obvious than at an impact parameter of 8 fm. Moreover, the results suggest that with an increase in the number of peripheral collisions, protons are more easily condensed with the magnetic field. Magnetic field effects are further investigated by the ratio of free neutrons to free protons as functions of a two-particle correlator $C_{2}$, four-particle correlator $C_{4}$ and six-particle correlator $C_{6}$ of angle $\phi$, rapidity $Y$ and transverse momentum $p_{T}$.  The results show that weak magnetic field effects could be revealed more clearly by these multiple-particle correlators, with the larger number of particle correlators demonstrating a clear signal. The results highlight a new method to search for weak signals using multi-particle correlators.  
\end{abstract}


\maketitle

\section{Introduction}
\label{introduction}

The Stern-Gerlach experiment is an important and renowned experiment, in which silver atoms are split due to an interaction between the spin magnetic moment and the magnetic field \cite{GS22}. Magnetic field effects are related to many aspects: the dynamics in nanofluids \cite{MS17}, the properties of neutron stars \cite{MD15,MB17}, as well as pion condensation \cite{LYZ18} \textit{etc}. Recently, magnetic field and their effects have been extensively  studied by the heavy ion collision community.  At ultra-relativistic energies, an extremely strong magnetic field up to $\approx$$10^{18}$ G can be created  \cite{VSK09,MAS10,ABZ12,WTD12,ZhaoXL_18_PRC,ChengYL_19}. With such an intense magnetic field, some anomalous transport phenomena could be induced in hot quantum chromodynamic (QCD) matter \cite{DK13,HXG16,HXG17NST,HXG20}, including the chiral magnetic effect (CME) \cite{KF08,BA13,LA14,VK17,HJX18,ZhaoXL2,WangFQ_NST}, chiral separation effect (CSE) \cite{DTS04}, chiral electric separation effect (CESE) \cite{HXG13,YJ15,YeYJ}, chiral magnetic waves (CMW) \cite{DEK11,ZhaoXL3,ShenDY}, chiral vortical effect (CVE) and global rotation \cite{DEK11-1,Deng19,XuZW_NST}. Conducted at the relativistic heavy ion collider (RHIC), results from the solenoidal tracker at RHIC (STAR) experiment show that $\Lambda$ particles are polarized in peripheral heavy ion collisions \cite{STAR17}, proposed by the global polarization mechanism of quark-gluon plasma (QGP) \cite{LZT05}. Such a global polarization phenomenon is relevant to vorticity fields as well as magnetic fields. A difference in polarization between $\Lambda$ and $\bar{\Lambda}$ was also observed in the STAR experimental results \cite{STAR17,STAR17B}. However, it remains a challenge to understand the differences between relativistic heavy ion collisions. One of the reasons could be due to extremely strong magnetic field itself; this issue is still hotly debated \cite{PLG17,ZhaoJ,FB17,HZZ18,GY19}. 
On the other hand, through the differences between $\Lambda$ and  $\bar{\Lambda}$ polarizations, one could derive the magnitude of the magnetic field \cite{PLG17}. However, reasonable magnetodynamics are not yet established for describing the evolution of QGP and hadronic matter. In addition, they are not fully understood for the hadronization process, which could affect the magnetic field as well as the system polarization. Due to intermediate-energy heavy ion collisions, the strength of the magnetic field is much less compared to the magnetic field strength in relativistic collisions \cite{LOU11}. Regardless, some magnetic field effects could also be investigated at this energy, such as pion production \cite{LOU11,GFW16}. Unfortunately, the work on exploring magnetic field effects in this energy domain is very limited. More effort should be focused towards this intermediate energy domain, and new methods or probes to detect tiny signals would be welcome. 

In our previous work \cite{DXG17,DXG18}, we focused on electromagnetic effects on the nucleon spectrum and photon production, using a one-body transport theory, the Boltzmann-Uehling-Uhlenbeck model, as a framework. Here, we have two purposes in this paper. One is to understand the nuclear interaction effect on the magnetic field during the collision process. The other purpose is to investigate the magnetic field effects on free nucleons by multiple-particle correlators.  

The rest of this paper is organized as follows. In Sec. \ref{TF1} we give a brief introduction of the isospin-dependent quantum molecular dynamics (IQMD) model. In Sec. \ref{TF2}, the nuclear interaction effect on magnetic field as well as multiple-particle correlators is investigated  via angular distributions. In our final section, we provide concluding remarks.

\section{NUMERICAL SETUP}
\label{TF1}
\subsection{Isospin-Dependent Quantum Molecular Dynamics model}

The quantum molecular dynamics (QMD) model, an $n$-body transport theory, has been developed for more than three decades \cite{JA91}. It contains information of emitted fragments, succeeding in reproducing various experimental data in nucleus-nucleus collisions \cite{XJ16,ZYX18,Feng_NST}. Naturally, the model can provide information about free nucleons (i.e. emitted protons and neutrons) during the evolution of the collisions. At present, the simulations are performed in the center of mass frame (CMS) within the IQMD model; an improved version of the QMD model, incorporating isospin-dependent  interactions and the Pauli exclusion principle. In this model, each nucleon is treated as a Gaussian wave packet in a coherent state \cite{JA91}:
\begin{eqnarray}
\phi_{i}(\vec{r},t) = \frac{1}{(2{\pi}L)^{3/4}}exp\Big{[}-\frac{(\vec{r}-\vec{r}_{i}(t))^{2}}{4L}+\frac{\textbf{i}\vec{p}_{i}(t)\cdot\vec{r}}{\hbar}\Big{]}, 
\label{QMDGaussWave}
\end{eqnarray}%
where $L$ is the square of the Gaussian wave packet widths and is fixed at 2.18 fm$^{2}$. Several interaction terms are included in the IQMD model as follows,
\begin{eqnarray}
V_{tot} = V_\text{sky} + V_\text{yuk} + V_\text{sym} + V_\text{MDI} + V_\text{Coul}, 
\label{QMDpotential}
\end{eqnarray}%
where the terms on the right-hand side correspond to Skyrme, Yukawa, symmetry, momentum-dependent, and Coulomb interactions, respectively.  The detailed forms of these interaction terms can be seen in Refs.~\cite{JA91,CH98}.

When considering a moving charged particle, one can appropriately implement an electromagnetic field into a transport model by  adding Li\'{e}nard-Wiechert potentials~\cite{LOU11,DXG18}. At position $\vec{r}$ and time $t$,
\begin{align}
&e\vec{B}(\vec{r},t) = \frac{e^2}{4\pi\epsilon_{0}c}\sum_{n}Z_{n}\frac{c^2-\upsilon^2_{n}}
{(cR_{n}-\vec{R}_{n}\cdot\vec{\upsilon}_{n})^3}\vec{\upsilon}_{n}\times\vec{R}_{n},
\label{eB11}
\end{align}
where the left-hand side is multiplied by a charge, $e$, in order to see the electromagnetic fine structure constant $\alpha = e^2/4\pi = 1/137$ (here $\epsilon_{0} = \hbar = c = 1$ ) appear on the right-hand side. Here, $Z_{n}$ is the charge number of the $n$-th particle, and $\vec{R}_{n} = \vec{r}-\vec{r}'_{n}$, where $\vec{r}'_{n}$ is the position of a charged particle moving with velocity $\vec{\upsilon}_{n}$ at retarded time $t_{rn} = t - |\vec{r}-\vec{r}'_{n}(t_{rn})|/c$. To simplify, only the magnetic field part is considered in our simulations, while the electric field contribution is replaced by a Coulomb interaction term. Considering this magnetic field, the Boltzmann equation reads as,
\begin{eqnarray}
\frac{\partial}{\partial t}{f} & + &\frac{\vec{p}}{m}\cdot\nabla_{\vec{r}}{f} +(- \nabla_{\vec{r}}U + F_\text{Lorentz})\cdot\nabla_{\vec{p}}{f} \notag\\
& = &\big{(}\frac{\partial {f}}{\partial t} \big{)}_{coll.},
\label{BUUequation}
\end{eqnarray}%
where the right-hand side is a collision term. The third term of Eq.~(\ref{BUUequation}) implies that the magnetic field will affect the distribution function, $f$.

In this work, one of our motivations is to investigate the effects of nuclear interaction on the magnetic field during the collision process; therefore, appropriate impact parameters must be considered. 
We choose two impact parameters. One is chosen large enough ($b$ = 11 fm) to reduce the projectile-target overlapping region and hence nuclear interaction, while the other has a moderate value ($b$ = 8 fm) so that suitable overlapping and nuclear interaction occurs. Furthermore, we define a $\phi$ angle by atan2($p_{x}$, $p_{z}$) in the $x-z$ reaction plane, since the Lorentz force of charged particles moving parallel to the $z$-axis is in the $x$-direction. 

\begin{figure}[htb]
\setlength{\abovecaptionskip}{0pt}
\setlength{\belowcaptionskip}{8pt}\centerline{\includegraphics[scale=0.56]{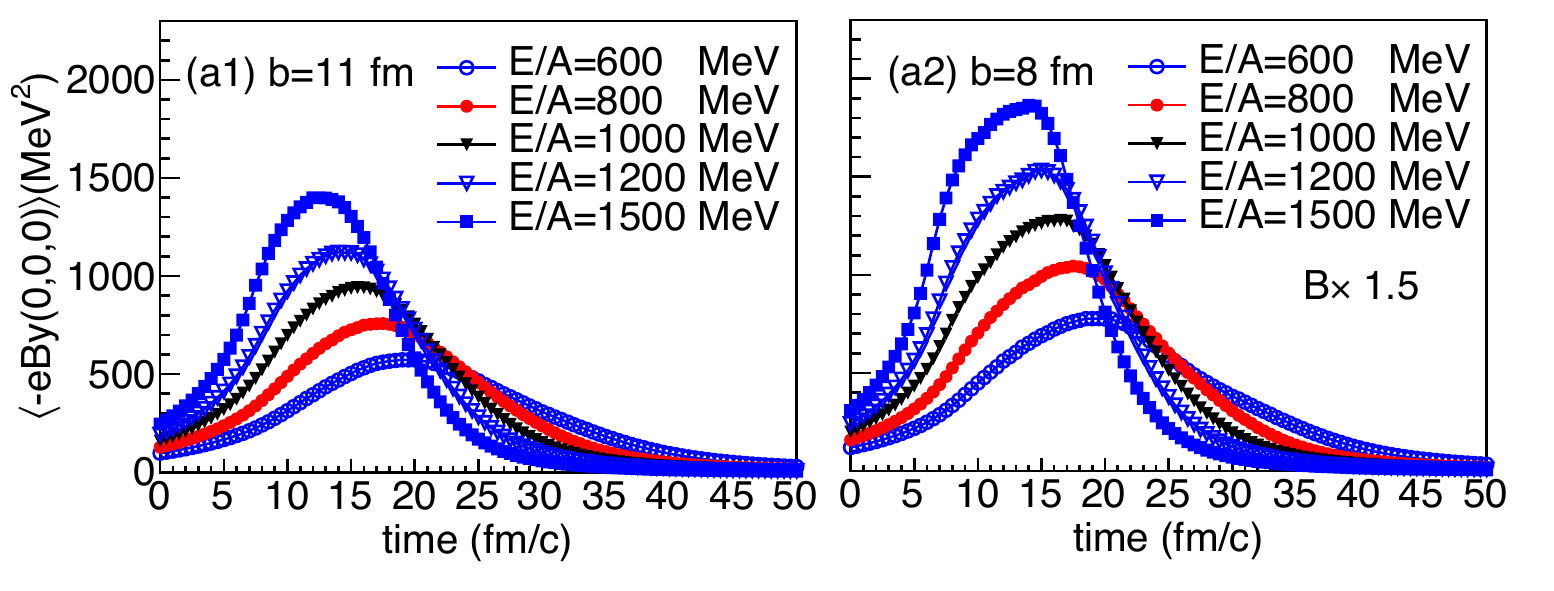}} 
\caption{Time evolution of magnetic field along $y$-axis at central point $R(0,0,0)$, with impact parameters of (a1) $b$ = 11 fm and (a2) $b$ = 8 fm, at different beam energies for $^{197}$Au + $^{197}$Au collisions. Note that for (a2), magnetic field is artificially multiplied by a factor of 1.5.}
\label{fig1}
\end{figure}

As mentioned previously, the two impact parameters of $b$ = 11 fm and $b$ = 8 fm were selected for comparison. These values were chosen to reduce the nuclear interaction effect as much as possible, while keeping some participant nucleons. Another reason for these specific values was to keep a similar strength of magnetic field for both impact parameters. However, as previous results \cite{LOU11,DXG17} have demonstrated, the strength of magnetic field ($B$) essentially increases with an increase in impact parameter. So, magnetic field at $b$ = 11 fm is higher than the magnetic field at $b$ = 8 fm. Therefore, to reduce the difference in magnetic field value, we artificially multiply a factor of 1.5 to amplify the original magnetic field for collisions at $b$ = 8 fm during simulations. After this procedure, we can see that the magnetic field given to collisions at $b$ = 8 fm is a little stronger than for $b$ = 11 fm. One can see the time evolution of the magnetic fields along the $y$-axis at central point $R(0,0,0)$ in Fig.~\ref{fig1}. Comparing Fig.~\ref{fig1} (a1) and (a2), the magnetic field at $b$ = 8 fm is stronger than the magnetic field at $b$ = 11 fm, due to the factor of 1.5 at the same beam energy. 

\begin{figure*}[htb]
\setlength{\abovecaptionskip}{0pt}
\setlength{\belowcaptionskip}{8pt}
\centerline{
\includegraphics[scale=1.0]{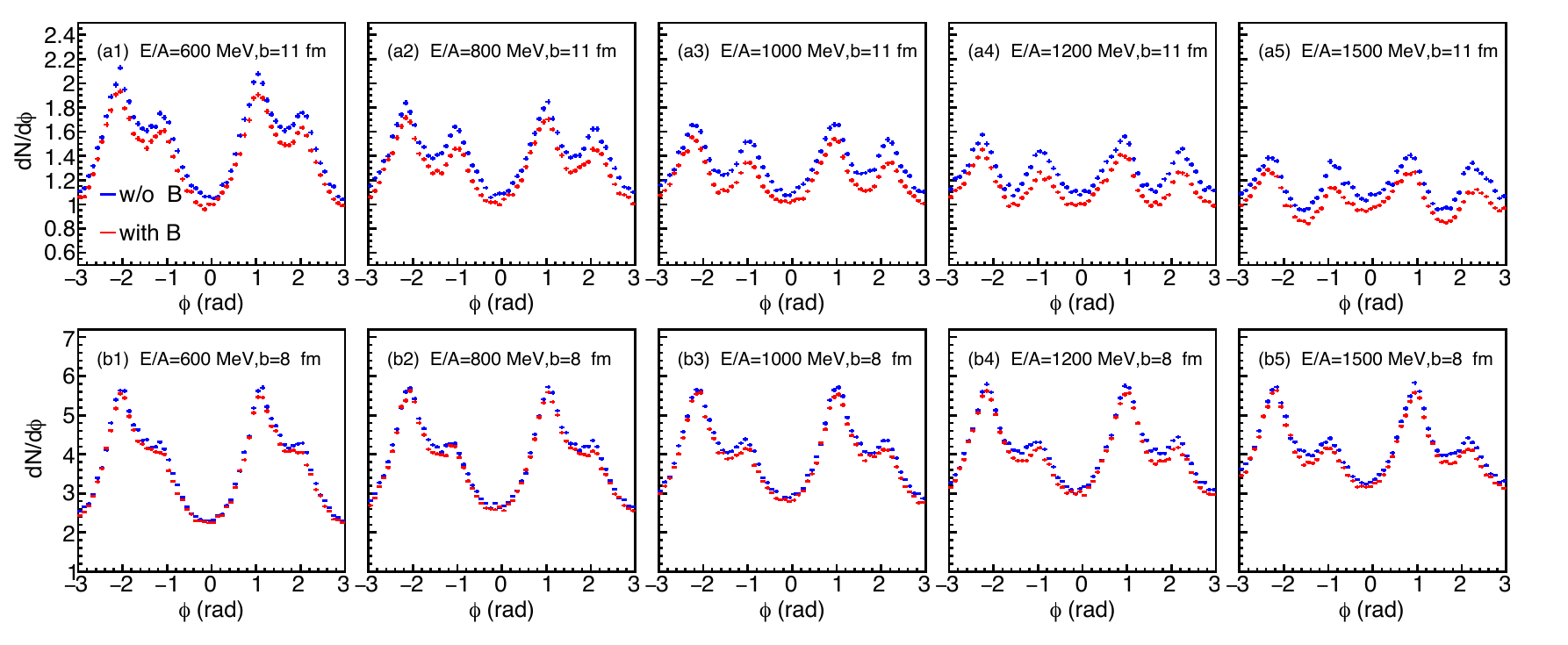}}
\caption{Distributions of proton angles, $\phi$, for $^{197}$Au + $^{197}$Au collisions at different incident energies and at impact parameters of $b$ = 11 fm (top panel) and $b$ = 8 fm (bottom panel), respectively. Red dots are results without $B$ and blue dots are results with $B$.}
\label{fig2}
\end{figure*}

\begin{figure*}[htb]
\setlength{\abovecaptionskip}{0pt}
\setlength{\belowcaptionskip}{8pt}
\centerline{
\includegraphics[scale=1.0]{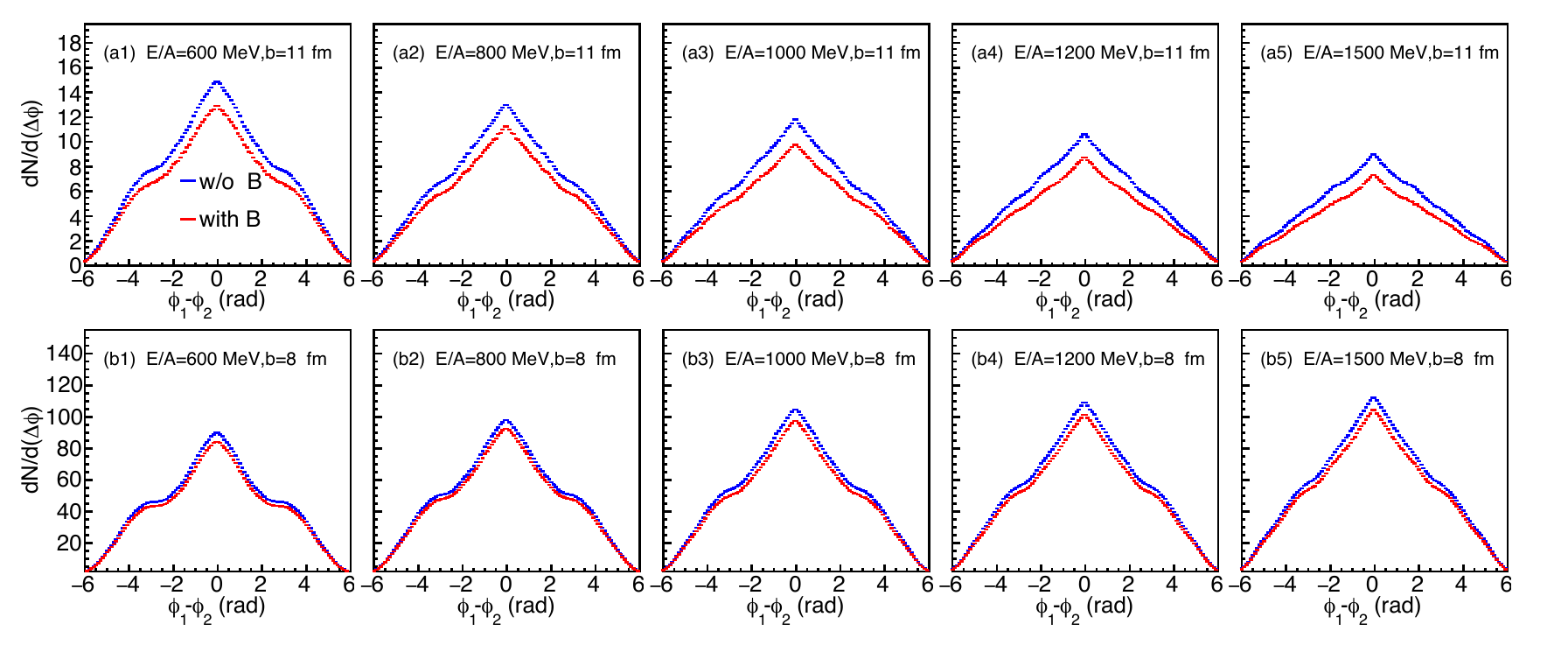}}
\caption{  Distributions of two-particle correlators of proton angles for $^{197}$Au + $^{197}$Au collisions at different incident energies and at impact parameters of $b$ = 11 fm (top panel) and $b$ = 8 fm (bottom panel), respectively. Red dots are results without $B$ and blue dots are results with $B$.}
\label{fig3}
\end{figure*}

In this work, the ratios of free neutrons to free protons \cite{LiBA02,Yan1,Yan2,LiWJ} as functions of angle $\phi$, rapidity $Y = \frac{1}{2}\frac{E+p_{z}}{E-p_{z}}$, and transverse momentum $p_{T} = \sqrt{p_{x}^{2}+p_{y}^{2}}$ are calculated. Here, we explain how to extract the $n/p$ ratio in more detail. One can get spectra of $\phi$, $Y$ and $p_{T}$ for free protons and free neutrons in the same event. When extracting the values in each bin of spectra of free protons and free neutrons and dividing correspondingly, we get the $n/p$ ratio in the same event. Through event-by-event accumulation, the average is calculated. In addition, we introduce multi-particle correlators $C_{n}$ between free nucleons, event by event, which are defined as:
\begin{eqnarray}
\label{C2CC}
&C_{2}&\approx Q_{1} - Q_{2}, \\
\label{C4CC}
&C_{4}&\approx Q_{1} + Q_{2} - Q_{3} - Q_{4}, \\
\label{C6CC}
&C_{6}&\approx Q_{1} + Q_{2} + Q_{3} - Q_{4} - Q_{5} - Q_{6},
\end{eqnarray}
where $Q$ represents either an angle $\phi$, rapidity $Y$, or transverse momentum $p_{T}$. The subscripts of $Q$ are particle indices (free protons or free neutrons) in the same event. For a two-particle correlator $C_{2}$, one can get $C_{2}$ distribution spectra of free protons with two loops of free protons in one event. In the same event and by the same approach, one can get $C_{2}$ distribution spectra of free neutrons. The $n/p$ ratio as a function of a two-particle correlator $C_{2}$ can subsequently be obtained. Furthermore, four-particle ($C_{4}$) and six-particle ($C_{6}$) correlators can get distribution spectra by the same method, but with four-particle loops and six-particle loops, respectively. Considering the symmetry, only even multi-particle correlators are adopted in this work.

\section{Results and discussion}
\label{TF2}
\subsection{Angular distributions}

First, distributions of proton angle $\phi$ are extracted at different beam energies and impact parameters, as displayed in Fig.~\ref{fig2}. It should be noted that the magnetic field effect is considered within the mid-rapidity region $|Y/Y_{0}|<0.5$, which is  dominant interaction zone. $Y_{0}$ is the initial rapidity of the projectile and is equal to $\frac{1}{2}\frac{E_\text{proj} + p_\text{proj,z}}{E_\text{proj} - p_\text{proj,z}}$. All of the distributions have four peaks, with each peak corresponding to the emission direction of a particle. The first and third peaks in Fig.~\ref{fig2} stem from a projectile and target passing through the third and first quadrants in the $x-z$ plane, respectively. The second and fourth peaks are formed around the angles $-\pi/4$ and $3\pi/4$, since particles squeeze-out by two nuclei in the fourth and second quadrants, respectively. It should be noticed that the values of the first and third peaks are higher than those of the second and fourth peaks, which indicates that these free protons mainly originate from spectators. 

One more interesting observed result is that the first and third peaks become smaller as the beam energy increases, in Fig.~\ref{fig2} (a1)–(a5). This means that spectators pass through more quickly and the duration time becomes shorter as beam energy increases. Thus, a few free protons would be emitted. However, it is the opposite for the smaller impact parameter in Fig.~\ref{fig2} (b1)-(b5). As incident energy increases, the structure of $\phi$ distributions do not drastically change, except for the quantities. What we are mostly concerned about here is the magnetic field effects. By comparing blue dots (no magnetic field present) and red dots (magnetic field present) in each panel of Fig.~\ref{fig2}, the magnetic field effects could be found, with fewer protons emitted  presenting for $b$ = 11 fm. This implies that protons could be condensed in the reaction region by the magnetic field \cite{GCY11}, similar to pion condensation in a magnetic field \cite{LYZ18}. However, for the case of $b$ = 8 fm, magnetic field effects are not so significant. This suggests that more peripheral collisions could generate the more obvious magnetic field effects, while protons would be easily condensed by the magnetic field.

Furthermore, angle differences between two protons at different beam energies and impact parameters are displayed in Fig.~\ref{fig3}. A difference between two cases with and without a magnetic field clearly manifests. The magnetic field effect can be seen more clearly by an angle correlator. This indicates that a tiny signal, originating from intermediate-energy heavy ion collisions, might be observed by using angle correlators. The magnetic field effect on free protons with an impact parameter of $b$ = 11 fm is more observable than with an impact parameter of $b$ = 8 fm. From previous results, one would have expected that for $b$ = 8 fm, with a stronger magnetic field as illustrated in Fig.~\ref{fig1}, we would see a stronger magnetic field effect on observables. However, this is not the case. Thus through the adapted Boltzmann equation~(\ref{BUUequation}), we believe that the main effect of the magnetic field could be washed-out by the nuclear interactions in the participant zone, including nucleon-nucleon collisions and the mean field term.

\begin{figure}[htb]
\setlength{\abovecaptionskip}{0pt}
\setlength{\belowcaptionskip}{8pt}\centerline{\includegraphics[scale=0.40]{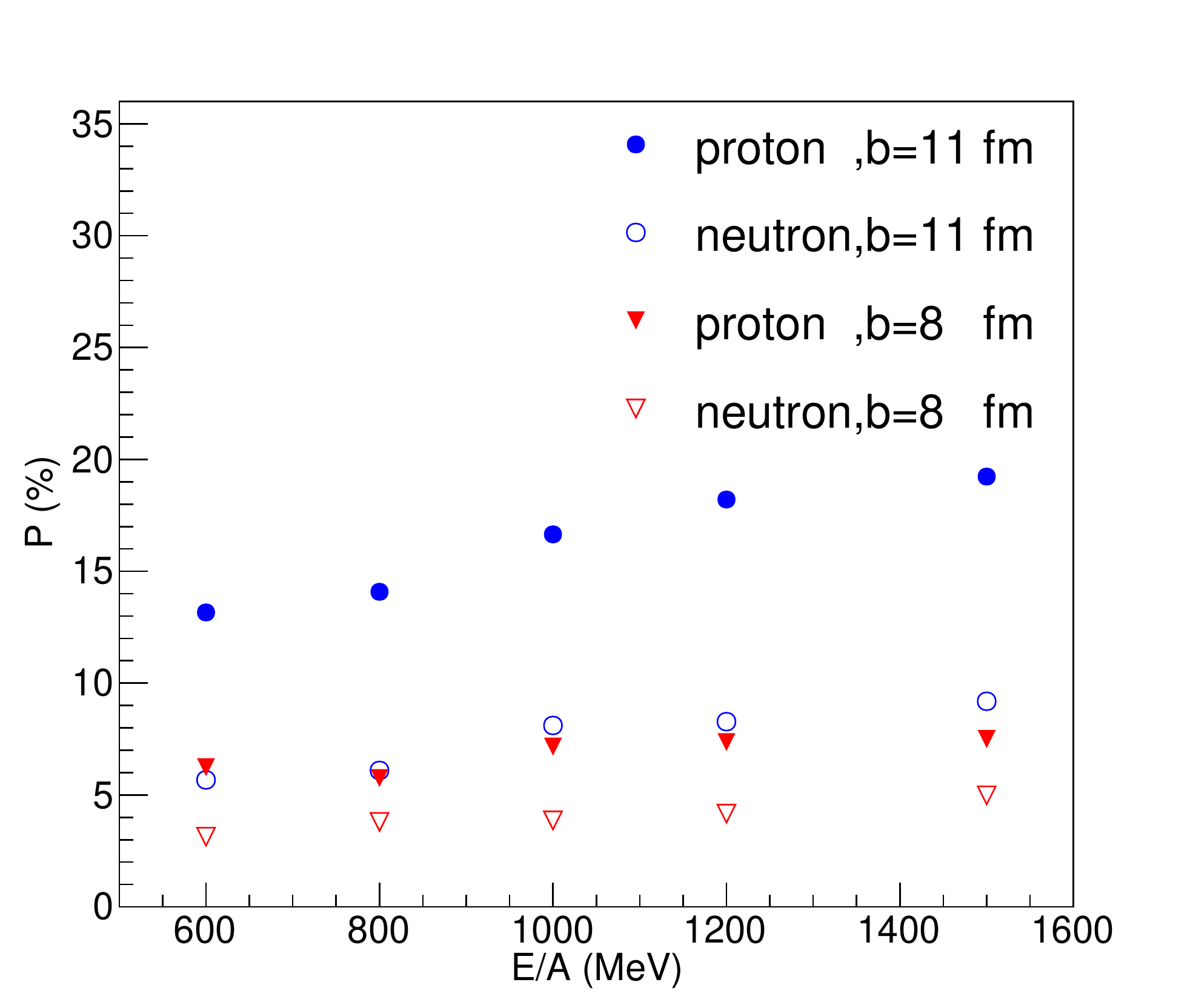}} 
\caption{ Ratio $P (\%)$ of average two-particle angle between the difference in cases with and without a magnetic field to the case without a magnetic field, for free protons and free neutrons at impact parameters $b$ = 11 fm and $b$ = 8 fm.}
\label{fig4}
\end{figure}

To discuss the incident energy dependence of this magnetic field effect, we consider a relative ratio $P$, which is defined by 
\begin{equation}
P = \langle \frac{ {A_{w/o}-A_{with}}}{ {A_{w/o}}} \rangle,
\label{eq_P}
\end{equation}
 where ``$A$'' is an index of value of an observable, eg. $dN/d(\Delta\phi)$, and $\langle \cdots \rangle$ denotes the average through each bin. We extract $P$ values by averaging over all points within the range of (-$\pi,\pi$) for $\phi$, or within (-1,1) for $Y$. Fig.~\ref{fig4} shows the $P$ values of $dN/d(\Delta\phi)$ a function of beam energy. Here, free neutrons are taken into account. Fig.~\ref{fig4} shows us that the ratio for free protons at $b$ = 11 fm reaches 20-percent, considerably larger than $P$ at $b$ = 8 fm. This indicates for the case of $b$ = 8 fm, that with a stronger magnetic field $B$ combined with stronger interactions from nucleon-nucleon collisions and the mean field interaction, it finally leads to smaller $B$ effect than for the case of $b$ = 11 fm. Thus, magnetic field effects are dampened by nuclear interaction. This suggests that a stronger magnetic field effect would be manifested in more peripheral collisions. Of course, the magnetic field not only affects protons, it also influences neutrons a little, which can be seen in  Fig.~\ref{fig4}. Nucleon-nucleon collisions and the mean field interaction can pass through interacting nucleons, thus affecting neutrons during the reaction process. However, these magnetic field effects on free neutrons are small compared to the effects on protons (Fig.~\ref{fig4}), due to the direct Lorentz force on protons. As shown in Fig.~\ref{fig4}, all values are positive; this means that both protons and neutrons could be trapped by the magnetic field.

\begin{figure}[htb]
\setlength{\abovecaptionskip}{0pt}
\setlength{\belowcaptionskip}{8pt}\centerline{\includegraphics[scale=0.49]{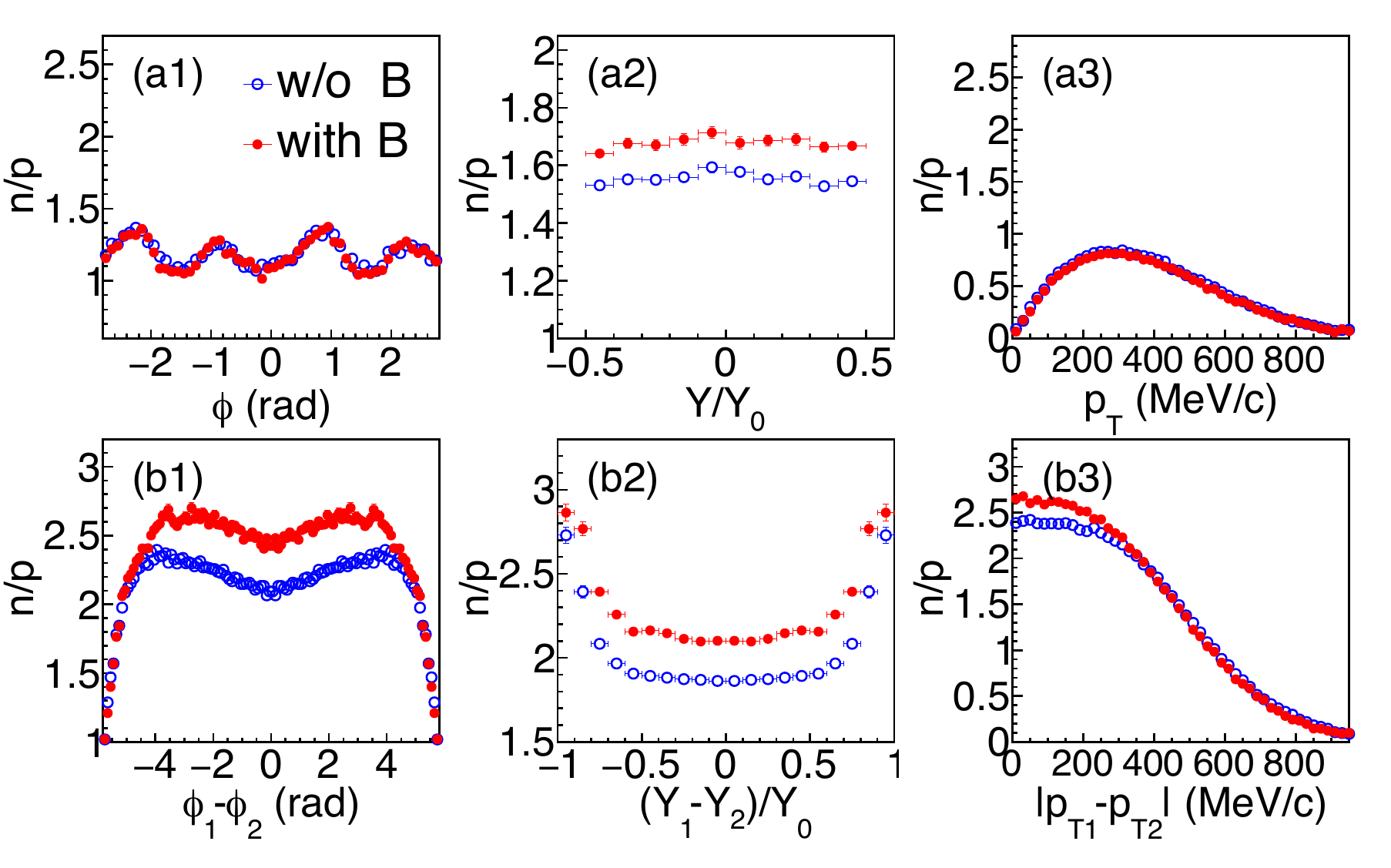}} 
\caption{ Top panels: $n/p$ ratios as functions of angle $\phi$, rapidity $Y/Y_{0}$, and transverse momentum $p_{T}$ without a magnetic field $B$ (blue circles), and with $B$ (red circle) for impact parameter $b$ = 11 fm and $E_{beam}/A$ = 1000 MeV. Bottom panels: $n/p$ ratio as functions of angle correlator $|\phi_{1}-\phi_{2}|$, rapidity correlator $(Y_{1}-Y_{2})/Y_{0}$, and transverse momentum correlator $|p_{T1}-p_{T2}|$  without $B$ (blue circles) and with $B$ (red circles) for impact parameter $b$ = 11 fm and $E_{beam}/A$ = 1000 MeV.}
\label{fig5}
\end{figure}

\begin{figure}[htb]
\setlength{\abovecaptionskip}{0pt}
\setlength{\belowcaptionskip}{8pt}\centerline{\includegraphics[scale=0.49]{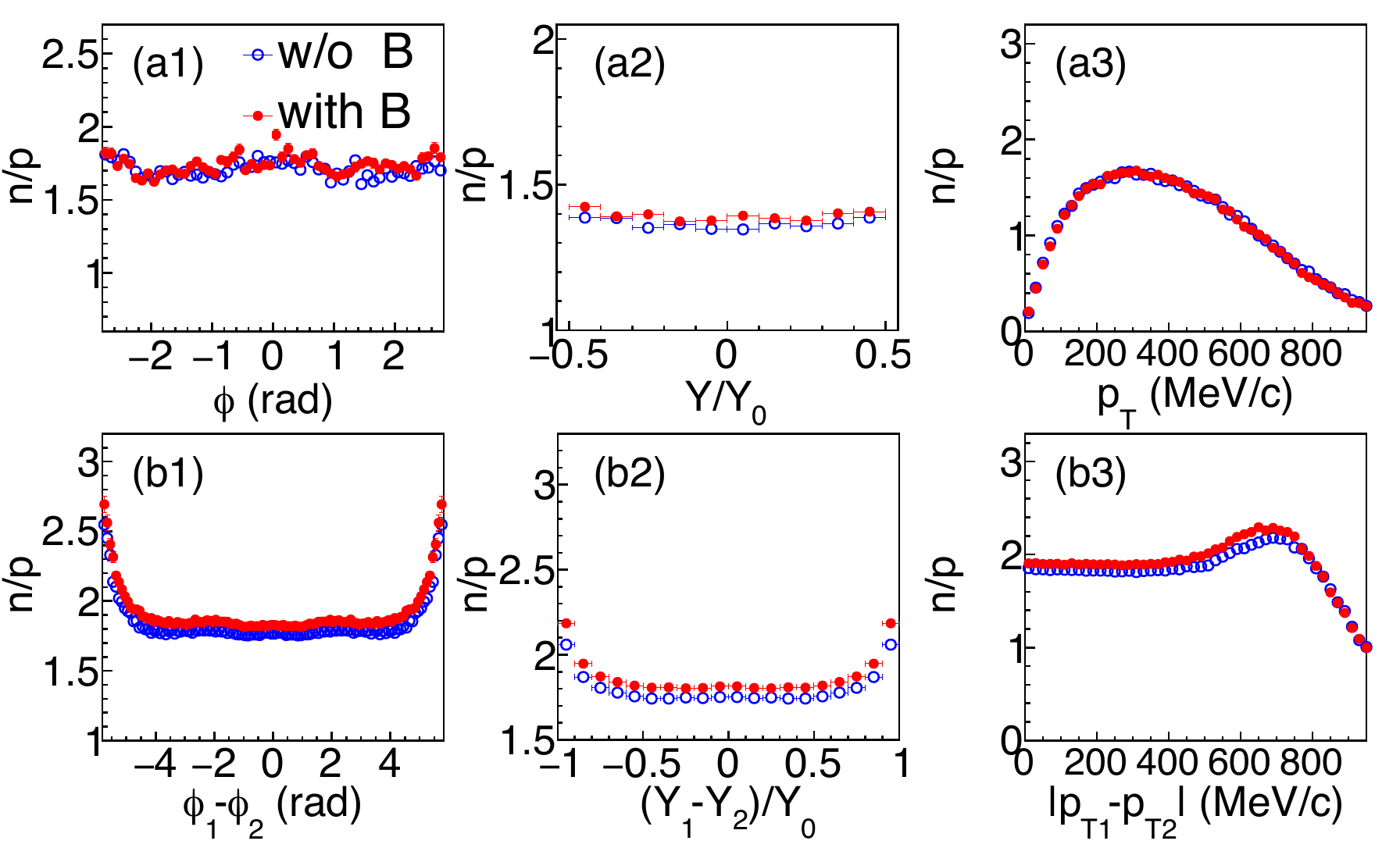}} 
\caption{ Top panels: $n/p$ ratios as functions of angle $\phi$, rapidity $Y/Y_{0}$, and transverse momentum $p_{T}$ without a magnetic field $B$ (blue circles), and with $B$ (red circle) for impact parameter $b$ = 8 fm and $E_{beam}/A$ = 1000 MeV. Bottom panels: $n/p$ ratio as functions of angle correlator $|\phi_{1}-\phi_{2}|$, rapidity correlator $(Y_{1}-Y_{2})/Y_{0}$, and transverse momentum correlator $|p_{T1}-p_{T2}|$  without $B$ (blue circles) and with $B$ (red circles) for impact parameter $b$ = 8 fm and $E_{beam}/A$ = 1000 MeV.}
\label{fig6}
\end{figure}

In addition to the angular distribution of correlators above, $n/p$ ratios as functions of angle correlator, rapidity correlator, and transverse momentum correlator are analyzed. As a comparison, the cases without correlators are shown in the top panels of Fig.~\ref{fig5}. As displayed in Fig.~\ref{fig5} (a1), at $b$ = 11 fm and 1000 MeV/nucleon,  the $n/p$ ratio displays a four-peak structure as a function of $\phi$. However, the magnetic field effect is not visible. A similar situation is seen for the $n/p$ ratio as a function of $p_{T}$ in Fig.~\ref{fig5} (a3), except a one-peak structure is displayed. For the $n/p$ ratio as a function of rapidity in Fig.~\ref{fig5} (a2), the presence of a magnetic field results in a higher $n/p$ ratio value, since less free protons are produced when $B$ is non-zero, as shown in Fig.~\ref{fig2}. Therefore, the rapidity-dependent $n/p$ ratio appears more sensitive to magnetic field than the angle and transverse momentum-dependent $n/p$ ratios. In the bottom panels of Fig.~\ref{fig5} in which particle correlations are considered, we found that magnetic field effects become much more obvious; there is even a distinguishable magnetic field effect on the $p_{T}$ correlator. The $|\phi_{1}-\phi_{2}|$ and $(Y_{1}-Y_{2})/Y_{0}$ correlators both exhibit clearer magnetic field effects. For the collisions at $b$ = 8 fm plotted in Fig.~\ref{fig6}, however, a significantly less visible magnetic field effect was discovered due to nuclear interactions (as discussed previously). 

\begin{figure}[htb]
\setlength{\abovecaptionskip}{0pt}
\setlength{\belowcaptionskip}{8pt}\centerline{\includegraphics[scale=0.49]{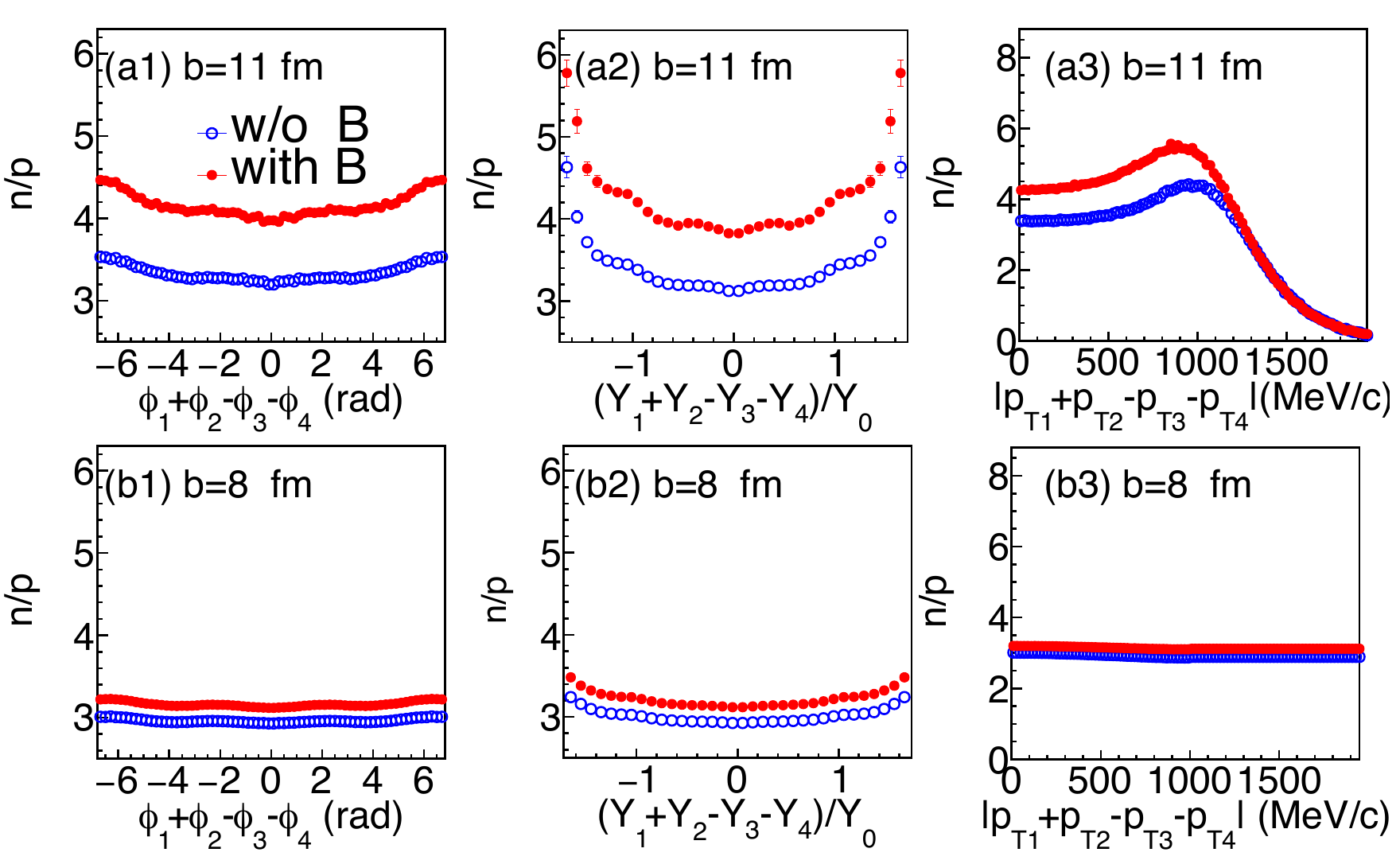}} 
\caption{$n/p$ ratios  as  functions of a four-particle angle correlator $\phi_{1}+\phi_{2}-\phi_{3}-\phi_{4}$, rapidity correlator $(Y_{1}+Y_{2}-Y_{3}-Y_{4})/Y_{0}$  and transverse momentum correlator $|p_{T1}+p_{T2}-p_{T3}-p_{T4}|$  without $B$ (blue circles) and with $B$ (red circles) at $b$ = 11 fm (top panels) and $b$ = 8 fm (bottom panels), at beam energy of 1000 MeV/nucleon.}
\label{fig7}
\end{figure}

\begin{figure}[htb]
\setlength{\abovecaptionskip}{0pt}
\setlength{\belowcaptionskip}{8pt}\centerline{\includegraphics[scale=0.40]{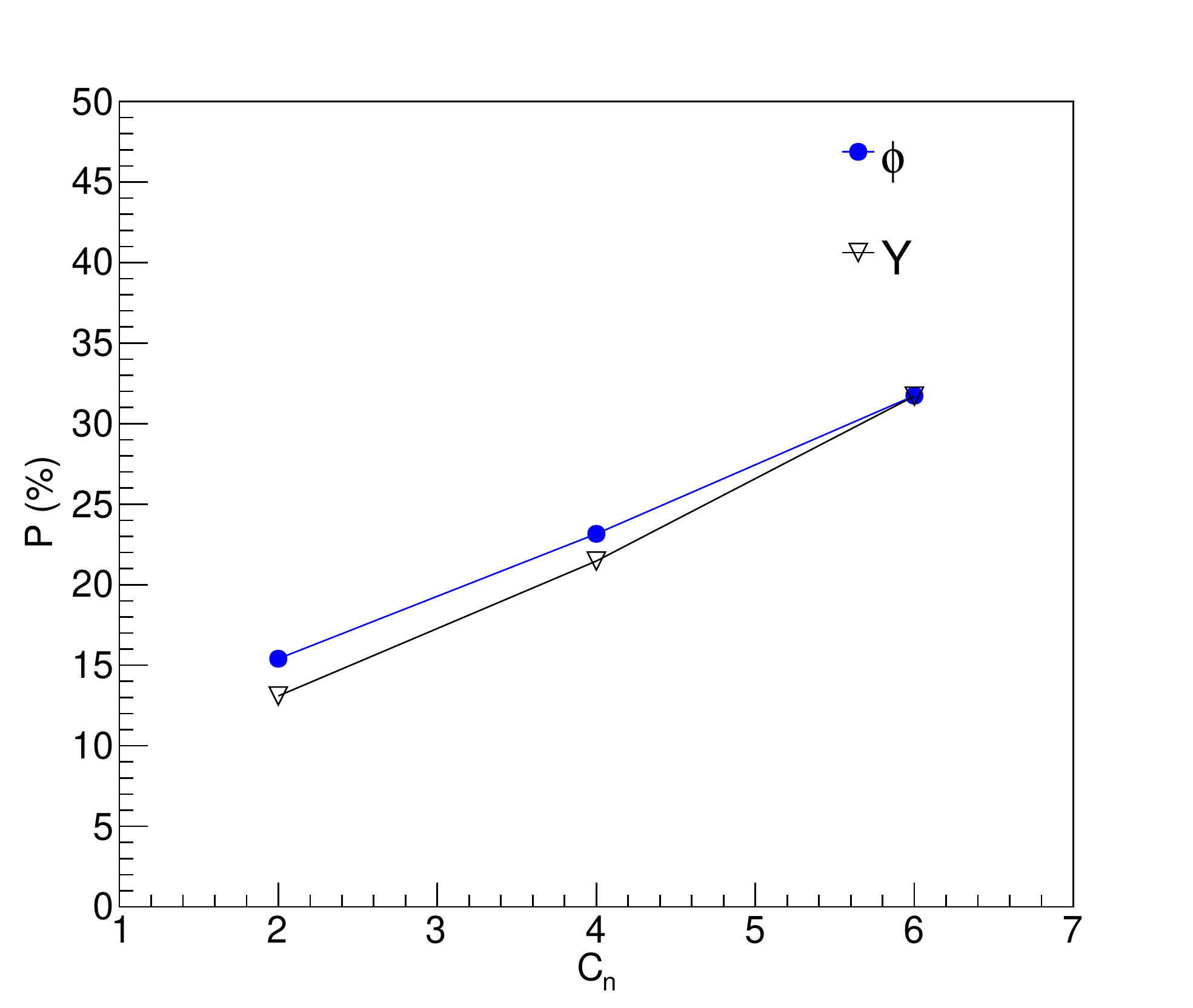}} 
\caption{Ratio  $P (\%)$ of differences from $n/p$ ratios between the cases without and with $B$ to the case without $B$ as a function of the $n$-particle correlator $C_{n}$ at  $b$ = 11 fm. The blue-solid circles represent data from angle $\phi$ and the black-empty triangles represent data from rapidity $Y$.} 
\label{fig8}
\end{figure}

As discussed previously, the magnetic field effect can be observed by a two-particle correlator. In general, can the same be said for multi-particle correlators? To answer this question, a four-particle correlator $C_{4}$ is analyzed. As seen from Fig.~\ref{fig7}, the multi-particle correlator demonstrates a larger magnetic field effect, even for the transverse momentum. With a change to a smaller impact parameter at $b$ = 8 fm, the values are still lower in comparison with the higher impact parameter values in the top panel, but the magnetic field effects are now visible. One might want to calculate the effects on a larger multi-particle correlator, such as a six-particle correlator $C_{6}$. However, calculations for a larger particle correlator would take up much CPU time, beyond our computational limit when calculating $C_{6}$ at $b$ = 8 fm. We still perform computations for the case at $b$ = 11 fm. Similar to Fig.~\ref{fig4}, using the definition of $P$ in Eq.~(\ref{eq_P}) for $n/p$ ratios, we extract the ratio $P (\%)$ of differences from $n/p$ ratios between the cases without and with a magnetic field to the case without a magnetic field as a function of the correlator $C_{n}$. The results for $C_{6}$ are only for the angle correlator and rapidity correlator, as shown in Fig.~\ref{fig8}. It is clear that for both angle and rapidity correlators, magnetic field effects are more obvious with an increase in $C_{n}$, and a relation of $P_{C_2}$$<$$P_{C_4}$$<$$P_{C_6}$ is discovered. The latter relationship illustrates that a larger multi-particle correlator is a more sensitive probe. Different from the behavior manifested in the top panels of Fig.~\ref{fig5}, i.e. the angle-dependent $n/p$ ratios are less sensitive than rapidity-dependent $n/p$ ratios, here they demonstrate similar sensitivity if using the correlators. This indicates that multi-particle correlators are a useful tool for revealing tiny signals, because information is superimposed when using multi-particle correlators.

\section{Conclusion}
\label{summary}

Magnetic field effects for $^{197}$Au + $^{197}$Au collisions, at beam energies ranging from 600 MeV/nucleon to 1500 MeV/nucleon, are investigated in the framework of the IQMD model. Initially, we investigated the magnetic field effect with distributions of proton angle $\phi$ and the two-particle correlator of angle $\phi$. To achieve this, a stronger magnetic field strength is artificially given to the  $^{197}$Au + $^{197}$Au collisions with an impact parameter of $b$ = 8 fm. However, it is found that the collisions with more participants and a stronger nuclear interaction at an impact parameter of $b$ = 8 fm display a weaker magnetic field effect than collisions with fewer participants and a weaker nuclear interaction at $b$ = 11 fm. This result indicates that the nuclear interaction causes magnetic field effects to reduce. By observing the production of free protons and neutrons, this result indicates that both nucleons could be condensed by the magnetic field, with more peripheral collisions, resulting in easier nucleon condensation. 
Moreover, we defined multiple-particle correlators $C(n)$ for angle, rapidity, and transverse momentum as new sensitive probes to reveal small magnetic field effects. By investigating the ratios of free neutrons to free protons as functions of the angle, rapidity, and transverse momentum correlators, magnetic field effects are displayed. Furthermore, it was found that magnetic field effects can be clearly seen by multi-particle correlators, with the larger the number of particle correlators, the more visible the magnetic field effects. Hence, this work highlights a new method to investigate small signals by multi-particle correlators. It is expected that this multi-particle correlator method can be used to explore other tiny effects, such as the chiral magnetic effect, equation of state, and nuclear $\alpha$-clustering etc.

\vspace{.5cm}
{\bf Acknowledgments---}  This work was partially supported by
the National Natural Science Foundation of China under Contract
Nos. 11890714, 11421505,  11947217 and 2018YFA0404404, China Postdoctoral Science Foundation Grant  No. 2019M661332, 
the Strategic Priority Research Program of the CAS under
Grant No. XDB34030200  and XDB16, and the Key Research Program of Frontier Science of CAS under Grant NO. QYZDJ-SSW-SLH002.

\end{CJK*}
\end{document}